\documentclass[aps,twocolumn,reprint,tightenlines,amsmath,
amssymb,nofootinbib,floatfix,superscriptaddress]{revtex4}
\usepackage{graphicx}
\usepackage{amsfonts}
\usepackage{color}
\usepackage{multibib}
\usepackage[colorlinks=true,urlcolor=blue,linkcolor=blue,citecolor=blue]{hyperref}
\usepackage[T1]{fontenc}
\usepackage{lmodern}
\usepackage{subfigure}
\usepackage{epstopdf}

\newcommand{\imp}{\affiliation{Institute of Modern Physics, Chinese Academy of
Sciences, Lanzhou 730000, China}}
\newcommand{\itp}{\affiliation{CAS Key Laboratory of Theoretical Physics, Institute of
Theoretical Physics,\\ Chinese Academy of Sciences, Beijing 100190,
China}}
\newcommand{\ucas}{\affiliation{School of Physical Sciences,
            University of Chinese Academy of Sciences,
            Beijing 100049, China}
}

\begin{document}

\title{Triangular singularity and a possible $\phi p$ resonance in the
$\Lambda^+_c \to \pi^0 \phi p$ decay}

\imp
\itp
\ucas

\author{Ju-Jun Xie}
\email{xiejujun@impcas.ac.cn}
\imp\ucas

\author{Feng-Kun Guo}
\email{fkguo@itp.ac.cn}
\itp\ucas

\date{\today}

\begin{abstract}

We study the $\Lambda^+_c \to \pi^0 \phi p$ decay by considering a triangle
singularity mechanism.
In this mechanism, the $\Lambda^+_c$ decays into the $K^* \Sigma^*(1385)$, the
$\Sigma^*(1385)$ decays into the $\pi^0 \Sigma$ (or $\Lambda$), and then the
$K^* \Sigma$ (or $\Lambda$) interact to produce the $\phi p$ in the final state.
This mechanism produces a peak structure around~$2020$~MeV. In addition, the
possibility that there is a hidden-strange pentaquark-like state is also
considered by taking into account the final state interactions of $K^* \Lambda$,
$K^* \Sigma$, and $\phi p$. We conclude that it is difficult to search for the
hidden-strange analogue of the $P_c$ states in this decay. However, we do expect
nontrivial behavior in the $\phi p$ invariant mass distribution. The predictions
can be tested by experiments such as BESIII, LHCb and Belle-II.

\end{abstract}

\maketitle

\bigskip

\section{Introduction}
\label{sec:introduction}

In 2015, two hidden-charm pentaquark-like structures, $P_c(4380)$
and $P_c(4450)$, were observed in the $J/\psi p$ invariant mass
spectrum via the $\Lambda^0_b \to K^- J/\psi p$ decay by the LHCb
Collaboration~\cite{Aaij:2015tga}. After they were observed, the two
$J/\psi p$ resonances were investigated within multiple theoretical
schemes with the aim to explain their nature (for more details and
references, see the recent reviews~\cite{Chen:2016qju,Guo:2017jvc}).
The existence of pentaquarks with hidden charm in that mass region
was already predicted in
Refs.~\cite{Wu:2010jy,Wu:2010vk,Wang:2011rga,Yang:2011wz,Xiao:2013yca}
by studying the interactions of anticharm mesons and charm baryons
using different models. Furthermore, it was pointed out in
Ref.~\cite{Guo:2015umn} (see also
Refs.~\cite{Liu:2015fea,Guo:2016bkl}) that a triangle singularity is
located very close to the $\chi_{c1}p$ threshold, $\simeq4.45$~GeV,
and thus at the $P_c(4450)$ mass. Such a singularity could produce a
narrow peak mimicking the behavior of a narrow resonance, which
requires the $\chi_{c1}$ and proton to rescatter in an $S$ wave into
the final state $J/\psi p$~\cite{Bayar:2016ftu}. This would
require quantum numbers $J^P = 1/2^+$ or $3/2^+$ for the peak. Notice that
although the $3/2^-$ and $5/2^+$ were reported as the most preferred quantum
numbers in the original LHCb publication~\cite{Aaij:2015tga}, $3/2^+$ remains
one of the favored possibilities in a later experimental analysis using an
extended model~\cite{Jurik:2016bdm}. Clearly, further investigations on the
$P_c(4380)$ and
$P_c(4450)$ structures, in particular from more processes and more
experiments, are needed. Since the $P_c$ structures were observed in
the decay mode $J/\psi p$, it is natural to expect that these
states, were they hadronic resonances, can be produced in
photo-production process $\gamma p \to P^+_c \to J/\psi p$ where
they will appear as $s$-channel
resonances~\cite{Wang:2015jsa,Kubarovsky:2016whd,Meziani:2016lhg,
Blin:2016dlf}.

Analogous to the hidden-charm pentaquark states, one may consider
the possible existence of hidden-strange pentaquarks $P_s$, in which
the $c\bar{c}$ pair is replaced by $s\bar{s}$. In fact, in the light
flavor sector below $2$ GeV, understanding the nature of the
$N^*(1535)$ resonance with spin parity $J^P = 1/2^-$ is very
challenging~\cite{Klempt:2007cp,Crede:2013sze}. One peculiar
property of the $N^*(1535)$ is that it couples strongly to the
channels with strangeness, such as the $\eta N$ and $K \Lambda$,
which is difficult to understand in the classical
three-constituent-quark models. This finds an explanation within the
chiral unitary approach in the work of Ref.~\cite{Inoue:2001ip}. The
strange decay properties of the $N^*(1535)$ resonance can also be
easily understood by considering large five-quark components in
it~\cite{Liu:2005pm,Helminen:2000jb,Zou:2007mk,An:2008xk}. Within
this pentaquark picture, the $N^*(1535)$ resonance could be the
lowest $L=1$ orbitally excited $uud$ state with a large admixture of
$[ud][us]\bar{s}$ pentaquark component. This makes the $N^*(1535)$
heavier than the $N^*(1440)$ and also gives a natural explanation of
its large couplings to the channels with
strangeness~\cite{Zou:2010tc}. In a very recent quark model
study~\cite{Gao:2017hya}, a $J^P = 1/2^-$ state with a mass varying
from $1873$ to $1881$ MeV is obtained, and its main component is
$\eta N$. This state could correspond to the resonance $N^*(1895)$
which has only an overall two-star status according to the review by
the Particle Data Group (PDG)~\cite{Olive:2016xmw}. Its existence is
supported by the analysis of the new $\eta$ photo-production
data~\cite{Kashevarov:2017kqb,Collins:2017sgu}, which finds that the
$N^*(1895)$ with $J^P =1/2^-$ is crucial in order to describe the
cusp observed in the $\eta$ photo-production at around $1896$~MeV as
well as the fast near-threshold rise of the total cross section of
the $\gamma p \to \eta' p$ reaction~\cite{Kashevarov:2017kqb}. In
Refs.~\cite{Kashevarov:2017kqb,Collins:2017sgu}, it was also pointed
out that the $N^*(1895)$ has strong couplings to both the $\eta N$
and $\eta' N$ channels.

At around $2$~GeV, a $\phi N$ bound state is predicted in several
models~\cite{Gao:2000az,Huang:2005gw,Gao:2017hya}. Such a $\phi N$
state can be viewed as a $P_s$ pentaquark. In
Ref.~\cite{Gao:2017hya}, a $J^P = 3/2^-$ state dominated by the
$\phi N$ component is obtained with a mass varying from $1949$ to
$1957$ MeV. Independently, a $J^P = 3/2^-$ $N^*$
resonance~\footnote{In the editions of the PDG review before 2012,
all the evidence for a $J^P = 3/2^-$ state with a mass above $1800$
MeV was filed under a two-star $N^*(2080)$. There is now
evidence~\cite{Anisovich:2011fc} of two states in this region, and
the PDG has associate the older data (according to masses) to two
states: a three-star $N^*(1875)$ and a two-star
$N^*(2120)$~\cite{Olive:2016xmw}.} with a mass about $2.1$~GeV is
proposed to explain the experimental
results~\cite{Kohri:2009xe,Kiswandhi:2010ub,Xie:2010yk,Kiswandhi:2011cq,
Kim:2011rm,He:2012ud,Xie:2013mua,Seraydaryan:2013ija,Dey:2014tfa,Dey:2014npa,
Kiswandhi:2016cav} on the associated strangeness production
reactions $\gamma p \to p \phi$, $\gamma p \to K^+ \Lambda(1520)$,
$\gamma p \to K^* \Lambda$ and $\gamma d \to d \phi$. The
forward-direction enhancement at around $W = 2.1$ GeV in the $\gamma
p \to p \phi$ reaction can be also reproduced by including a special
correlated five-quark configuration of a color-antitriplet ($su$)
diquark and a color-triplet $[\bar{s}(ud)]$, which subsequently
hadronize into the $\phi$ and
proton~\cite{Lebed:2015fpa,Lebed:2015dca}. However, it is pointed
that such a five-quark configuration is not literally a resonant
pentaquark state~\cite{Lebed:2015fpa,Lebed:2015dca}. In
Ref.~\cite{He:2017aps}, it is proposed that the $J^P=3/2^-$ states
$N^*(1875)$ and $N^*(2100)$ in the $\phi$ photo-production are
hadronic molecular states from the $\Sigma^* K$ and $\Sigma K^*$
interactions, respectively, and they can be regarded as the
hidden-strange partners of the LHCb pentaquarks.

The $P_c$ structures were produced in the process $\Lambda_b^0\to
K^- J/\psi p$. Analogously, one may study the possible $P_s$ states
in the singly Cabibbo suppressed process $\Lambda_c^+\to \pi^0 \phi
p$. As pointed out in Ref.~\cite{Lebed:2015dca}, the $\Lambda^+_c
\to \pi^0 P^+_s \to \pi^0 \phi p$ and $\Lambda^0_b \to K^- P^+_c \to
K^- J/\psi p$ are entirely comparable if one substitutes
$V_{cb}V^*_{cs} \to V^*_{cs}V_{us}$. One important difference
between the two processes is that the former has a much smaller
phase space---the $\Lambda_c^+$ is above the $\pi^0\phi p$
three-body threshold only by 193~MeV. Such a small phase space for
the $\Lambda^+_c$ decay restricts that only the neutral pion is
possible in the final state, for a hadronic decay, if we want to
produce in addition a $p\phi$ pair. Because of the small phase space
and the weak $\pi^0\phi$ interaction\footnote{The $\pi^0\phi$
interaction should be very weak for two reasons. Firstly, the pion
and the $\phi$ meson do not have the same quark flavors, which leads
to an Okubo--Zweig--Iizuka (OZI) suppression. Secondly, the small
phase space means that the pion is soft, and the interaction between
a soft pion and matter fields is weak because of the spontaneous
breaking of chiral symmetry in quantum chromodynamics.}, no other
resonances except for the possible $P_s$ contribute to the process.
The first experimental measurement of the $\Lambda^+_c\to \pi^0 \phi
p$ process has been reported by the Belle
Collaboration~\cite{Pal:2017ypp}. Very recently, the Belle
Collaboration reported their searching for the decay of $\Lambda^+_c
\to \pi^0 p \phi$, and no significant signal was observed with an
upper limit on the branching fraction of ${\cal B}(\Lambda^+_c \to
\pi^0 p \phi) < 15.3 \times 10^{-5}$ at a $90\%$ confidence
level~\cite{Pal:2017ypp}.

\begin{figure*}[tbh]
\begin{center}
\includegraphics[width=\textwidth]{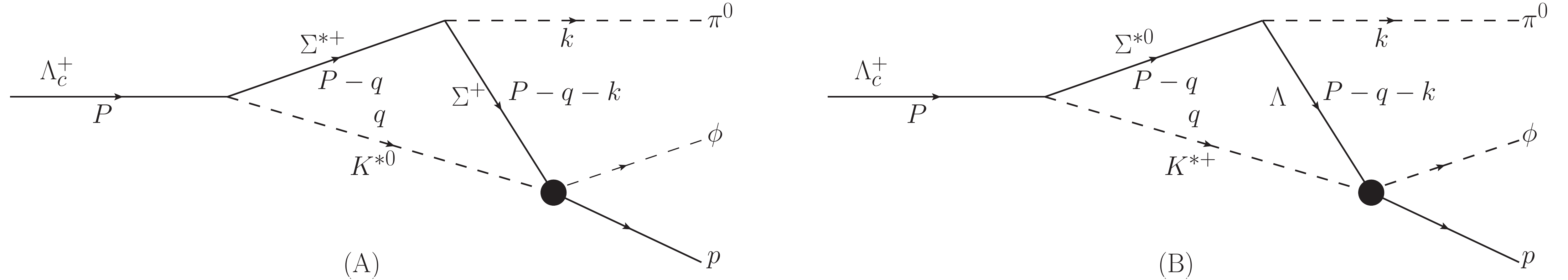}
\caption{Triangle diagrams for the $\Lambda^+_c \to
\pi^0 p \phi$ decay. ($A$): $\Sigma^+$-exchange. ($B$):
$\Lambda$-exchange. The definitions of the kinematical variables
($P, q, k$) are also shown.}
\label{fig:feydiagram}%
\end{center}
\end{figure*}

In this paper, we will show that the $\Lambda^+_c\to \pi^0 \phi p$ also receives
a contribution from triangle singularities close to the physical region.
A triangle singularity appears on the physical boundary in a
particular situation when all the intermediate states are on shell,
and all the particles move along the same direction (parallel or
anti-parallel) such that the interactions at all three vertices can
happen as classical processes~\cite{Coleman:1965xm}. Such a physical
picture can be easily seen following the analysis of
Ref.~\cite{Bayar:2016ftu}. In addition to the works related to the
$P_c$ structures mentioned above, the role played by triangle
singularities has been broadly investigated recently in the
literature~\cite{Wu:2011yx,Aceti:2012dj,Wu:2012pg,Liu:2014spa,
Szczepaniak:2015eza, Ketzer:2015tqa,Liu:2015taa,Szczepaniak:2015hya,
Aceti:2016yeb,Yang:2016sws,Wang:2016dtb,
Xie:2016lvs,Debastiani:2016xgg,
Roca:2017bvy,Debastiani:2017dlz,Samart:2017scf,Sakai:2017hpg}. Along
this line, we will calculate the triangle singularity contribution
to the $\Lambda^+_c \to \pi^0 \phi p$ decay, where the $\Lambda^+_c$
decays into $K^* \Sigma^*(1385)$, the $\Sigma^*(1385)$ ($\equiv
\Sigma^*$) decays to the $\pi^0 \Sigma$ (or $\Lambda$) and the $K^*
\Sigma$ (or $\Lambda$) rescatter into $\phi p$ in the final state,
see Fig.~\ref{fig:feydiagram}. In addition to the effects of the
triangle mechanism, we consider also the final state interaction
(FSI) of $K^* \Lambda \to \phi p$ and $K^* \Sigma \to \phi p$. Were
there a $P_s$ resonance, it must couple to both the $\phi p$ and
$K^*\Sigma/\Lambda$ and thus may be manifest in the Dalitz plot or
in the $\phi p$ invariant mass distribution. Yet, because of the
small phase space and depending on the mass and width of such a
$P_s$ state, it could be difficult to search for it. As will be
shown in this paper, on one hand the triangle singularity
contribution can enhance the production of such a resonance, on the
other hand it makes the identification of the $P_s$ signal more
difficult if its mass is around 2.02~GeV.

This paper is organized as follows. In Sec.~\ref{sec:formalism}, we
discuss the triangle diagrams and how a $P_s$ is included in our
model. The numerical results are presented in
Sec.~\ref{sec:results}, and finally a short summary is given in
Sec.~\ref{sec:summary}.

\section{Formalism}\label{sec:formalism}

The decay $\Lambda^+_c \to \pi^0 p \phi$ can proceed through the
triangle diagrams depicted in
Fig.~\ref{fig:feydiagram}\footnote{Replacing the $\Lambda$ by the
$\Sigma^0$ leads to vanishing contribution because the $\Sigma^{*0}$
cannot couple to the $\Sigma^0\pi^0$.}. Given the masses of the
initial state $\Lambda_c$, the neutral pion in the final state and
two of the intermediate states, for example the $K^*$ and
$\Sigma/\Lambda$, the region for the $\Sigma^*$ mass in order to
produce a triangle singularity at the physical boundary, i.e., in
the physical region\footnote{The triangle singularity of course
cannot be exactly in the physical region since otherwise one would
get a logarithmically divergent amplitude. It is shifted into the
complex plane because of the finite decay width of at least one of
the intermediate states. The amplitude in the physical region is
well defined without any divergence.} can be worked
out~\cite{Guo:2015umn,Liu:2015taa}. Using the central values for all
of the mentioned hadron masses, the region can be obtained as
$[1386.6, 1390.1]$~MeV for diagram~(A),  while the measured mass of
the $\Sigma^{*+}$, $(1382.80\pm0.35)$~MeV, is 4~MeV below. The
region is  $[1384.8, 1394.3]$~MeV for the $\Lambda$-exchange in
diagram~(B), while the measured mass of the $\Sigma^{*0}$
$(1383.7\pm1.0)$~MeV almost reaches the lower bound. In this case,
the triangle singularities still have sizeable influence on the
physical decay amplitude. The lower bound of that region means that
the triangle singularity in the $\phi p$ invariant mass is located
exactly at the two-body threshold of the two particles which
rescatter into the $\phi p$. Thus, one expects that the triangle
singularity induced effects would be mainly because of the
$\Lambda$-exchange diagram which can produce an enhancement around
the $K^{*+}\Lambda$ threshold at 2007~MeV. In the following, we give
the details of the calculation which shows explicitly the
enhancement around that energy.


The decay of $\Lambda^+_c \to (\Sigma^* K^*)^+$ can proceed by $W$-exchange
diagram~\cite{Chau:1995gk,Lu:2009cm,Cheng:2010vk}:
$(cd)u \to (du)u$, and the $duu$ are hadronized, together with a $s\bar{s}$ pair
with the vacuum quantum numbers, into the $\Sigma^*K^*$.

The evaluation of the diagram in in Fig.~\ref{fig:feydiagram} requires us first
to provide an expression for the $\Lambda^+_c \to (\Sigma^* K^*)^+$ vertex.
Because the $\Sigma^* K^*$ threshold ($\simeq 2277$ MeV) is very close to
the mass of $\Lambda^+_c$, we consider only the $S$-wave coupling. Then we can
write
\begin{eqnarray}
t_{\Lambda^+_c \to \Sigma^* K^*} &=& f_I g_{\Lambda_c \Sigma^* K^*}
\bar{u}^{\mu}(P-q) u(P) \varepsilon_{\mu}(q), \label{Eq:vertex1}
\end{eqnarray}
where $f_I$ is the isospin factor with $f_I = \sqrt{2/3}$ for the
$\Sigma^{*+} K^{*0}$ and $- \sqrt{1/3}$ for the $\Sigma^{*0} K^{*+}$, and
$g_{\Lambda_c \Sigma^* K^*}$ is an effective coupling constant which
can be obtained, in general, from the branching ratio of $\Lambda^+_c \to
\Sigma^* K^*$.

The decays of $\Sigma^* \to \pi \Sigma$ and $\pi \Lambda$ are in $P$
waves, then we can easily write with $SU(3)$ symmetry
\begin{eqnarray}
t_{\Sigma^{*+} \to \pi^0 \Sigma^+} &=& \frac{g}{m_{\pi}}
\bar{u}(P-q-k)u^{\mu}(P-q)k_{\mu}, \\
t_{\Sigma^{*0} \to \pi^0 \Lambda} &=& \sqrt{3} \frac{g}{m_{\pi}}
\bar{u}(P-q-k)u^{\mu}(P-q)k_{\mu},
\end{eqnarray}
with $g = 0.69$ obtained from the total decay width
$\Gamma_{\Sigma^*} = 37.13$~MeV and the branching fraction ${\rm
Br}[\Sigma^* \to \pi \Sigma] = 0.117$.

After the production of the $K^{*0} \Sigma^+$ and $K^{*+} \Lambda$, they
rescatter into the $\phi p$ in the final state, as shown in
Fig.~\ref{fig:feydiagram}. The total decay amplitude for the processes shown in
Fig.~\ref{fig:feydiagram} can be written as
\begin{eqnarray}
t  &\!\! = &\!\!  \frac{g_{\Lambda_c \Sigma^* K^*} \, g}{m_{\pi}}
\vec{\epsilon}_\phi \cdot \vec{k} \sum_{i = \Sigma,\Lambda}
\mathcal{C}_{i} \int\frac{{\rm d}^4q}{(2\pi)^4} \nonumber \\
&\!\!&\!\! \times
\frac{i2m_{\Sigma^*}}{(P-q)^2-m^2_{\Sigma^*}+im_{\Sigma^*}\Gamma_{\Sigma^*}}
\frac{i}{q^2 - m^2_{K^*} + im_{K^*}\Gamma_{K^*}} \nonumber \\
&\!\!&\!\! \times \frac{i2m_i}{(P-q-k)^2 - m^2_i + i\epsilon},
\end{eqnarray}
where we have defined $\mathcal{C}_{\Sigma} = \frac{\sqrt{6}}{3} \,
t_{K^{*0} \Sigma^+ \to \phi p}$ and $\mathcal{C}_{\Lambda} = -
\,t_{K^{*+} \Lambda \to \phi p}$, and $t_{K^{*0} \Sigma^+ \to \phi
p}$ and $t_{K^{*+} \Lambda \to \phi p}$ are $T$-matrix elements for
the rescattering processes, which will be discussed in the next section. We
notice that the $K^* \Sigma^*$ mass threshold is close to the mass of
$\Lambda^+_c$ and the range of the $\phi p$ invariant mass for the decay of
interest, $[1957.7, 2141.5]$~MeV, allows us to make nonrelativistic
apporoximation for all the involved baryons and vector mesons. Therefore, we can
consider only $S$ waves for the rescattering. Furthermore, we can
make the approximation
\begin{eqnarray}
\sum|\vec{\epsilon}_\phi \cdot \vec{k}| \simeq |\vec{k}|^2,
\label{Eq:sumpolarization}
\end{eqnarray}
where the sum runs over the polarizations of the $\phi$ meson.

After performing the contour integration over the temporal component $q^0$ in
Eq.~\eqref{Eq:totalamplitude}, in the same way as shown in
Refs.~\cite{Bayar:2016ftu,Aceti:2015zva}, and including the finite widths of the
$\Sigma^*$ and $K^*$ resonances, we get
\begin{eqnarray}
&& t  =  - \frac{g_{\Lambda_c \Sigma^* K^*} \, g}{m_{\pi}}
\vec{\epsilon}_\phi \cdot \vec{k} \, m_{\Sigma^*} \, t_T, \label{Eq:totalamplitude} \\
&& t_T =  \sum_{i = \Sigma,\Lambda} \mathcal{C}_{i} \, m_i \int
\frac{{\rm d}^3q}{(2\pi)^3} \frac{1}{\omega_{K^*} E_{\Sigma^*} E_i}
\nonumber \\
&& \times \frac{1}{k^0 - E_i - E_{\Sigma^*} +
i\,{\Gamma_{\Sigma^*}}/{2}} \frac{1}{P^0 + \omega_{K^*} + E_i -
k^0} \nonumber
\\
&& \times \frac{1}{P^0 - \omega_{K^*} - E_i - k^0 + i\,
{\Gamma_{K^*}}/{2}} \nonumber \\
&& \times \frac{\left [ P^0 \omega_{K^*} + k^0 E_i - (\omega_{K^* +
E_i})(\omega_{K^*} + E_i + E_{\Sigma^*}) \right ] }{P^0 - E_{\Sigma^*} - \omega_{K^*}
+ i\, {\Gamma_{\Sigma^*}}/{2}} , \nonumber
\label{Eq:totalamplitude-3dimention}
\end{eqnarray}
where $\omega_{K^*} = \sqrt{m^2_{K^*} + |\vec{q}|^2}$, $E_{\Sigma^*}
= \sqrt{m^2_{\Sigma^*} + |\vec{q}|^2}$, $P^0 = M_{\Lambda^+_c}$,
$k^0 = \sqrt{m^2_{\pi^0} + |\vec{k}|^2} = (M^2_{\Lambda^+_c} +
m^2_{\pi^0} - M^2_{\phi p})/(2M_{\Lambda^+_c})$, and $E_i =
\sqrt{m^2_i + |\vec{q} + \vec{k}|^2}$ with $i = \Sigma$ or
$\Lambda$. Because the $S$-wave vertices attached to the $\Lambda_c$
initial state and the $\phi p$ final state do not introduce any momentum
dependence into the loop amplitude, and the $P$-wave pionic vertices result in
a factor of the pion momentum, the above loop integral is ultraviolet
convergent.

The $\phi p$ invariant mass mass distribution for the $\Lambda^+_c \to \pi^0
\phi p$ decay then reads
\begin{eqnarray}
\frac{d\Gamma}{dM_{\phi p}} &=& \frac{m_p m^2_{\Sigma^*}
g^2_{\Lambda_c \Sigma^* K^*} \, g^2}{16\pi^3 M_{\Lambda^+_c}
m^2_\pi} |\vec{k}|^3 |\vec{p}_{\phi}| |t_T|^2, \label{Eq:dgam}
\end{eqnarray}
where $\vec{k}$ is the $\pi^0$ momentum in the rest frame of the
$\Lambda^+_c$, and $\vec{p}_{\phi}$ is the $\phi$ momentum in the
center-of-mass frame of the $\phi p$ system. They are given by
\begin{eqnarray}
|\vec{k}| \!\! &=& \!\! \frac{\sqrt{[M^2_{\Lambda^+_c} - (m_{\pi^0} + M_{\phi p})^2][M^2_{\Lambda^+_c} - (m_{\pi^0} - M_{\phi p})^2]}}{2M_{\Lambda^+_c}}, \nonumber \\
|\vec{p}_{\phi}| \!\! &=& \!\! \frac{\sqrt{[M^2_{\phi p} - (m_{\phi}
+ m_{p})^2][M^2_{\phi p} - (m_{\phi} - m_{p})^2]}}{2M_{\phi p}},
\nonumber
\end{eqnarray}
with $m_{\pi^0} = 134.98$~MeV, $m_\phi = 1019.46$~MeV, and $m_p = 939.27$~MeV.
Finally, the partial decay width of the
$\Lambda^+_c \to \pi^0 p \phi$ decay is obtained by integrating
Eq.~\eqref{Eq:dgam} over $M_{\phi p}$,
\begin{eqnarray}
\Gamma = \int^{M_{\Lambda^+_c} - m_{\pi^0}}_{m_\phi + m_p} dM_{\phi p} \,
\frac{d\Gamma}{dM_{\phi p}} \, .  \label{Eq:Gamma}
\end{eqnarray}

\section{Numerical results} \label{sec:results}

So far we have not specified the input for the rescattering
$T$-matrix elements. In principle, because of the very small phase
space and the closeness of the thresholds, all of the involved
hadrons, $K^*$, $\Sigma$ or $\Lambda$, $\phi$ and proton, can be
treated nonrelativistically. Thus, one may construct a
nonrelativistic effective field theory describing the interaction
between vector mesons and baryons with the leading order defined by
a few constant contact terms. However, it does not make much sense
doing it in that manner because of the lack of experimental
information. We will thus take the model of Ref.~\cite{Oset:2009vf}
where the interaction of the vector mesons with the SU(3) octet
baryons is studied in the local hidden gauge formalism using a
coupled-channel unitary approach. In that model, a degenerate pair
of resonances with $J^P=1/2^-$ and $3/2^-$ which couple strongly to
$K^*\Sigma$, $K^*\Lambda$ and $\phi p$ is obtained, and the pole is
at $(1977+i\,55)$~MeV~\cite{Oset:2009vf}. They can be regarded as
the $P_s$ states. The prediction in this model was updated in light
of the $\gamma p\to K^0\Sigma^+$ data~\cite{Ewald:2011gw} in
Ref.~\cite{Ramos:2013wua} to get resonance parameters with a mass of
$2035$~MeV and a width of $125$~MeV. However, this state only shows
up in the transitions involving the $K^* \Sigma$ channel. One may
regard that model as providing a special set of parameters for the
nonrelativistic effective field theory mentioned above. By adjusting
the interaction strengths, one can in principle investigate the
possibility of $P_s$ with other masses and as well as the
possibility without any $P_s$, i.e. no pole around the $\phi p$
threshold.

\begin{figure}[tb]
\hglue-3mm\includegraphics[scale=0.46]{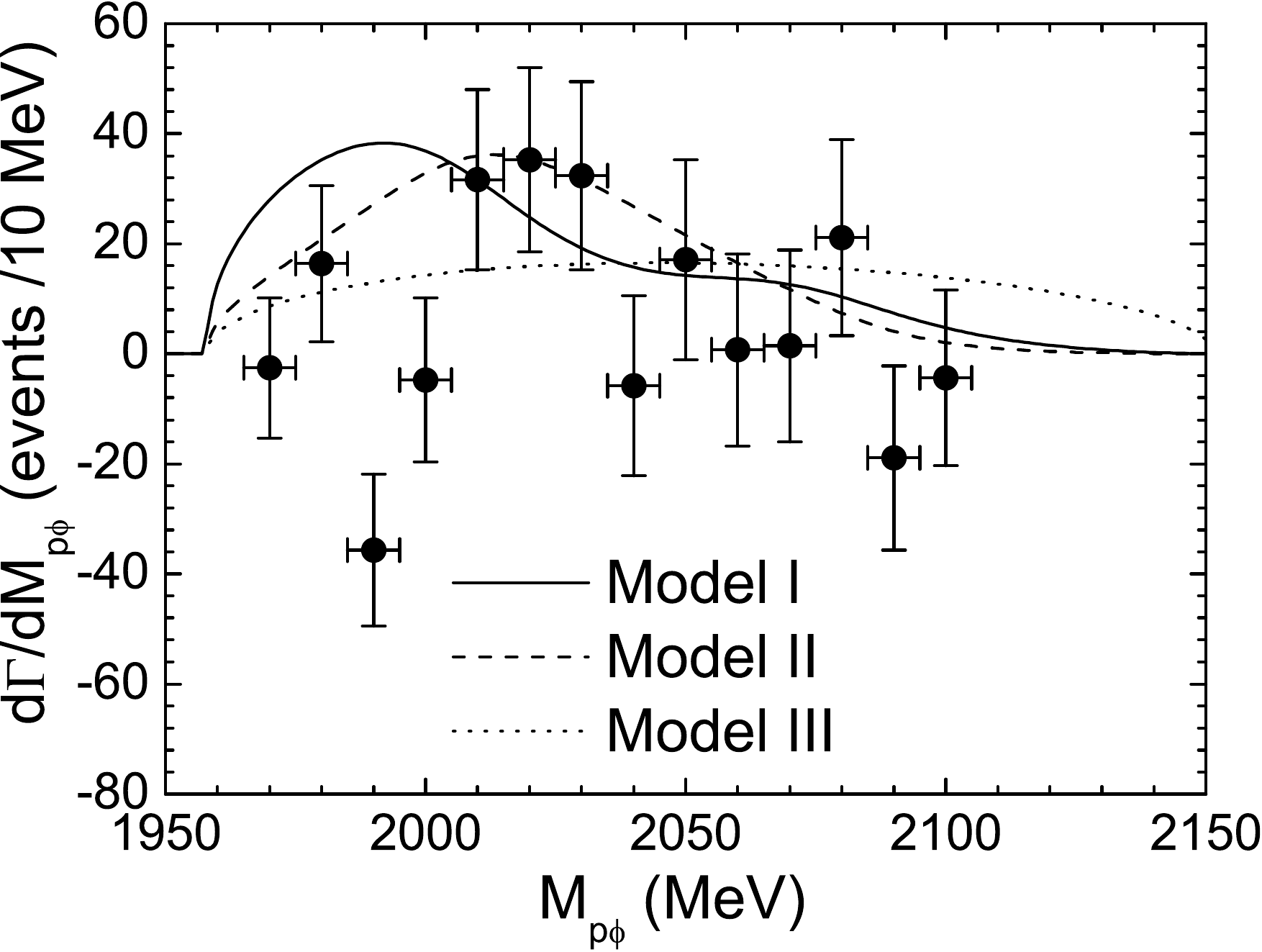}
\caption{Invariant mass distribution of the $\Lambda^+_c \to \pi^0 p
\phi$ decay. The experimental data are taken from Ref.~\cite{Pal:2017ypp}.}
\label{fig:dgdm}
\end{figure}

Here we present the numerical results for the $\phi p$ invariant
mass distribution for three different cases, which are denoted as
Model~I, II and III, in Fig.~\ref{fig:dgdm}. Model I represents the
calculation of the triangle diagrams in Fig.~\ref{fig:feydiagram}
with the FSI taken from Ref.~\cite{Ramos:2013wua}, which includes
the contribution of a $P_s$ state with properties specified in that
model, which could be well different in other models and in reality,
see above. Model II is different from Model I by modelling the FSI
by a constant, and it thus represents the case without any $P_s$
resonance. For comparison, we show the phase space without any
special dynamics as Model III. To be more explicit, for these three
cases, the total decay amplitudes $t_j$ ($j = {\rm I}$, II, and III)
are given by
\begin{eqnarray}
t_{\rm I} &=& t , \nonumber \\
t_{\rm II} &=& t , ~{\rm but ~ with ~}t_{K^{*+} \Lambda \to \phi p} =
\frac{\sqrt{6}}{2} t_{K^{*0}\Sigma^+ \to \phi p} \nonumber \\
&=& c_1 \frac{\sqrt{6}}{2} \frac{E_{K^*} + E_\phi}{4 F^2_\pi},
\label{Eq:tmodelII} \\
t_{\rm III} &=& c_2,
\end{eqnarray}
where $t$ is the amplitude shown in Eq.~\eqref{Eq:totalamplitude}, $c_1$ and
$c_2$ are normalization constants to be adjusted to match the measured event
distribution, $F_\pi=92.2$~MeV is the pion decay constant, $E_{K^*}$ and
$E_{\phi}$ are the energies of the $K^{*+}$ and $\phi$ mesons in the $\phi p$
center-of-mass frame. Here we take $E_{K^*} = 891.66$~MeV and $E_{\phi} =
1043.26$~MeV, which are obtained at the $K^{*+} \Lambda$ mass threshold.

In Fig.~\ref{fig:dgdm}, the solid, dashed, and dotted curves represent the
results of Model I, II, and III, respectively. The parameter $c_1$ of Model II
has been adjusted to the strength of the experimental data reported by the Belle
Collaboration~\cite{Pal:2017ypp} at its peak around $M_{\phi p} = 2020$~MeV. The
results of Model I and III are normalized such as to have the same integrated
partial width as Model II. Model II clearly shows a peak structure around
2.02~GeV. The origin of this peak is the triangle diagrams, in particular the
$\Lambda$-exchange in Fig.~\ref{fig:feydiagram}~(B) which has a triangle
singularity close to the $K^*\Lambda$ threshold ($\simeq 2.01$~GeV). The width
of this peak is comparable with the width of the $K^*$, which is about 50~MeV.
This is a quite natural consequence as the $\phi p$ invariant mass is the same
as the $K^*\Lambda$ invariant mass so that its distribution inherits the width
of the $K^*$. Were the $K^*$ width much smaller, one would get a much narrower
peak. For Model I, one might think that there should be also a bump structure
around $2035$~MeV which is the mass of the generated resonance in the vector
meson--baryon interaction model we are using~\cite{Ramos:2013wua}.
However, the triangle diagram involving the $K^* \Lambda \to \phi p$ transition
is the predominant contribution in the present case because its triangle
singularity is closer to the physical region, while the resonance peak only
shows up in the channels involving the $K^*\Sigma$.
\begin{figure}[tb] \hglue-3mm\includegraphics[scale=0.46]{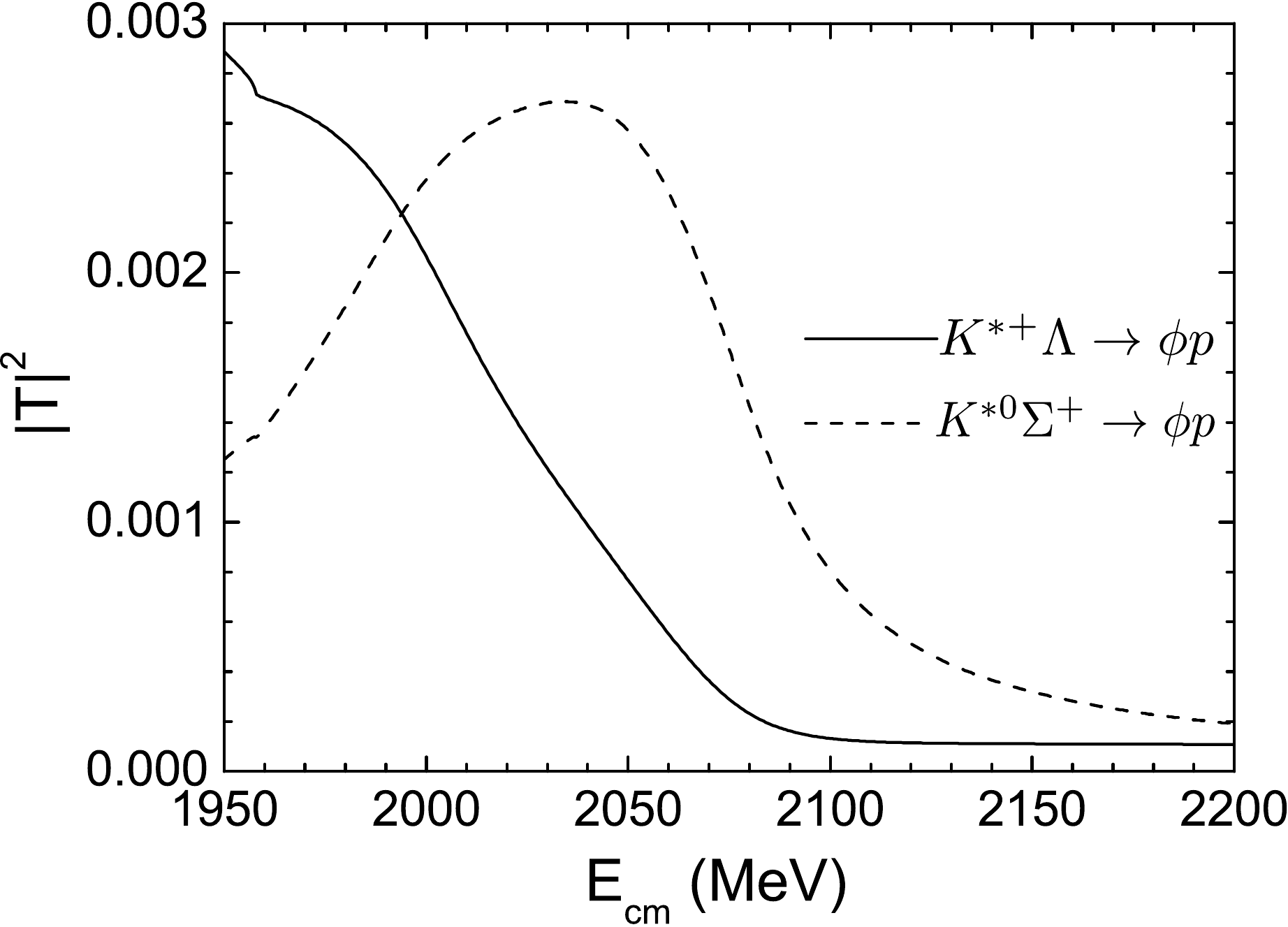} \caption{The
squared norm of the $T$-matrix elements for $K^{*+}\Lambda\to \phi p$ and
$K^{*0}\Sigma^+\to \phi p$ as a function of the meson--baryon invariant mass
$E_{\rm cm}$ in the model of Ref.~\cite{Ramos:2013wua}.}
\label{fig:Tsq}
\end{figure}
Here, the FSI results in a near-threshold enhancement, see Fig.~\ref{fig:Tsq}
where the kink in the solid line is located at the $\phi p$ threshold. It could
be that in other models the resonance couples to these vector meson--baryon
channels in a different pattern so as to show up as a near-threshold peak in the
$\phi p$ invariant mass distribution. Identifying such an enhancement in
experiments is difficult as it requires the data to have a high statistics. In
particular, it becomes much more difficult if the $P_s$ mass is close to the
$K^*\Lambda$ threshold because of the presence of kinematic singularities there.
In any case, the phase space shown as Model~III is very different from both
Model~I and Model~II. Despite the low statistics of the current Belle data, the
curve of Model~II, whose shape is completely fixed, has a remarkable agreement
with the data.
In particular, the data seem to indeed have a peak around $K^*\Lambda$
threshold. More data are welcome to clarify the situation.

Using the value of $g_{\Lambda_c \Sigma^* K^*}$ estimated in
Appendix~\ref{app}, we can get an estimate of the branching fraction
of the three-body decay $\Lambda_c^+\to\pi^0\phi p$ by integrating
over the $\phi p$ invariant mass distribution. For Model II with
$c_1 =1$, the result is
\begin{eqnarray}
{\rm Br}(\Lambda_c^+\to\pi^0\phi p)_{\rm II} =
\mathcal{O}\left(10^{-4}\right),
\end{eqnarray}
which is of the same order as the Belle upper limit~\cite{Pal:2017ypp}.
Thus, although not all of the contributions to the $\Lambda^+_c \to
\pi^0 \phi p$ decay are from this mechanism we expect the actual branching
fraction is of this order.

\section{Summary} \label{sec:summary}

We study the $\Lambda^+_c \to \pi^0 \phi p$ decay by considering a
triangle singularity mechanism. The decay was proposed to be a
channel to search for the hidden-strange partner of the $P_c$
states. The mechanism is such that the $\Lambda^+_c$ decays into the
$K^* \Sigma^*$, the $\Sigma^*$ subsequently decays into the $\pi^0
\Sigma$ (or $\Lambda$), and the $K^* $ then interacts with the
$\Sigma$ (or $\Lambda$) to produce the $\phi p$ in the final state.
In the $K^*\Sigma/\Lambda \to \phi p$ FSI, we consider cases with
and without a $P_s$ state. For the case with the $P_s$, we take the
model of Refs.~\cite{Oset:2009vf,Ramos:2013wua} which produces a
resonance at around 2~GeV. The triangle singularities considered in
this paper are close to the physical region, and can produce a peak
at around 2.02~GeV with a width similar to that of the $K^*$
resonance. The obtained $\phi p$ invariant mass distribution agrees
with the existing Belle data. Were there a $P_s$ state, it could
distort the distribution. However, it is difficult to be identified
in the decay under study because of the small phase space and the
presence of triangle singularities. We look forward to more data
from the BESIII, Belle-II and LHCb experiments in the future, which
will be decisive to illuminate the role played by triangle
singularities in this decay.

\section*{Acknowledgments}

We would like to thank Eulogio Oset and Wei Wang for useful discussions. This
work is partly supported by the National Natural Science Foundation of China
(NSFC) under Grant Nos.~11475227 and 11647601, by the DFG and NSFC through
founds provided to the Sino-German CRC 110 ``Symmetries and the Emergence of
Structure in QCD'' (NSFC Grant No.~11621131001), by the CAS Key Research Program
of Frontier Sciences (Grant No.~QYZDB-SSW-SYS013),  by the Youth Innovation
Promotion Association of CAS (Grant No.~2016367), and by the Thousand Talents
Plan for Young Professionals.

\bigskip

\begin{appendix}

 \section{Estimate of the $\Lambda_c\Sigma^*K^*$ coupling constant}
 \label{app}

The branching fraction for $\Lambda^+_c \to \Sigma^* K^*$ has not been estimated
so far. Yet, an upper limit has been reported as ${\rm Br}(\Lambda^+_c \to
\Lambda K^+ \pi^+\pi^-)<5\times10^{-4}$~\cite{Olive:2016xmw}. Because the
$\Sigma^*$ and $K^*$ decay dominantly into the $\Lambda\pi$ and $K\pi$, we thus
take $10^{-4}$ as an order-of-magnitude estimate for $\Lambda^+_c \to
\Sigma^* K^*)$ to estimate the coupling constant $g_{\Lambda_c \Sigma^* K^*}$ using the following
decay width formula
\begin{eqnarray}
&& \Gamma[\Lambda^+_c \to \Sigma^{*}K^*] = \frac{g^2_{\Lambda_c
\Sigma^* K^*}|\vec{p_1}|}{48\pi}\left(1+ \frac{m_{\Sigma^*} +
m_{K^*}}{M_{\Lambda^+_c}}\right) \times  \nonumber \\
&& \!\!\! \!\!\! \left(1 + \frac{m_{\Sigma^*} - m_{K^*}}{M_{\Lambda^+_c}}\right)
\left ( 8 + \frac{(M^2_{\Lambda^+_c} - m^2_{\Sigma^*} -
m^2_{K^*})^2}{m^2_{\Sigma^*} m^2_{K^*}} \right ),
\end{eqnarray}
with
\begin{eqnarray}
|\vec{p}_1| \!\! &=& \!\! \frac{\sqrt{[M^2_{\Lambda^+_c} -
(m_{\Sigma^*} + m_{K^*})^2][M^2_{\Lambda^+_c} - (m_{\Sigma^*} -
m_{K^*})^2]}}{2M_{\Lambda^+_c}}. \nonumber
\end{eqnarray}
Using the measured masses $M_{\Lambda^+_c} = 2286.46$ MeV,
$m_{\Sigma^*} = 1384.57$ MeV, $m_{K^*} = 893.1$ MeV and the total
decay width of $\Gamma_{\Lambda^+_c} = 3.29 \times 10^{-9}$ MeV, we
get
\begin{eqnarray}
g_{\Lambda_c \Sigma^* K^*} \sim 2 \times 10^{-7}.
\label{eq:g}
\end{eqnarray}

\end{appendix}

\end{document}